\documentclass[twocolumn,secnumarabic,amssymb, amsmath, nofootinbib,tightenlines,
nobibnotes, aps, prb]{revtex4}
\pdfoutput=1 
\usepackage{graphicx}% Include figure files
\usepackage{dcolumn}% Align table columns on decimal point
\usepackage{bm}% bold math

\begin{document}

\title{Opposite Changes in Gap Width of Opposite Spin States Induced by Rashba Effect in Anti-ferromagnetic Graphene on Ni(111) }% Force line breaks with \\

\author{C.H.Jin$^1$}
\author{Z.Bi$^1$}
\author{Z.H.Hu$^1$}
\author{J.Feng$^2$}
\author{E.G.Wang$^{1,2}$}
\affiliation{$^1$ School of Physics, Peking University, Beijing, China\\
$^2$ International Center of Quantum Materials (ICQM), Peking University, Beijing, China}
\date{\today}

\begin{abstract}
Graphene is a promising candidate for applications in spintronics. In this paper, Density Functional Theory method is used to calculate the band structure and magnetic properties of graphene on Ni(111). Our results show that once there is anti-ferromagnetic order in graphene, an external electric field at the order of $10^9$ V/m can induce a gap width difference of tens of meV for opposite spin states near the Fermi surface. 
\end{abstract}

\pacs{PACS:}% PACS, the Physics and Astronomy
                             % Classification Scheme.

\maketitle

\section*{\uppercase\expandafter{\romannumeral1}.\ INTRODUCTION}
Graphene is a two-dimensional material with many interesting properties\cite{28,29,30,31}. Behavior of Dirac Fermions in graphene has been extensively studied. Particularly, magnetic properties have attracted lots of interests.

Graphene is considered a promising candidates for  spintronics due to its long spin relaxation time and large spin relaxation length\cite{10}. Spin splitting is usually a desirable effect in potential applications.  Although the intrinsic spin orbit coupling (SOC) is quite weak\cite{20}, considerable extrinsic SOC effects have been proposed and studied\cite{11,12,13,14,15,16,17,18,19,20}, among which the Rashba SOC resulted from broken inversion symmetry received lots of attention. The Rashba Hamiltonian can be expressed as

\begin{equation}
H=\alpha(\mathbf{p}\times\mathbf{\sigma})\cdot \mathbf{E}
\end{equation}

$\alpha$\ ,\ $\mathbf{p}=\hbar \mathbf{k}$\ ,\ $\sigma$ and $\mathbf{E}$
respectively stands for Rashba strength parameter, electron momentum, Pauli spin operator and the electric filed. 

Graphene/Ni system is studied here for two reasons: One is that graphene films can be conveniently prepared on Ni substrate; the other is that Ni is a common candidate for spin injection electrode and spin detection electrode. Published work suggests that the Rashba-induced spin gap can be as large as tens to hundreds of meV\cite{9,17,18}.

However, systematic study of the influences of Rashba effect on the Dirac fermions around the K point near the Fermi surface which greatly affect the transport properties is still lacking.

In this paper, we study few layer graphene(FLG)/Ni(111) system using density-functional theory (DFT) by considering various factors including substrate interaction, layer number and SOC.

\section*{\uppercase\expandafter{\romannumeral2}.\ METHODS}

\begin{figure}[]
\includegraphics[width=7cm]{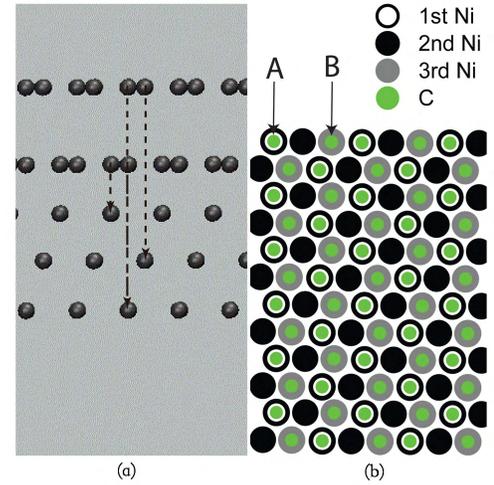}
\caption{\label{f1}(Color online) (a) Side view of bilayer graphene/Ni(111) system. Three lower layers and two upper ones respectively stand for Ni and graphene. (b) Top view of top-fcc structure.}
\end{figure}

For First Principle Calculations LDA with PAW in VASP\cite{24,25} has been used to optimize the Few Layer Graphene (FLG)/Ni system, which according to previous works is effective and sufficient\cite{21}. The energy cutoff of optimization is 500eV and the convergence criteria is that the Hellmann-Feynman forces be less than 0.03 eV/A for the geometry optimization. The graphene layers and  upper three layers of the total six layers Ni(111) are allowed to relax. A vacuum slab more than $15{\AA}$ is applied to avoid interaction between supercells.

The original $1\times1$ unit cell of Ni substrate is used, onto which the $1\times1$ unit cell of graphene is mapped. In monolayer case the graphene is of the well studied top-fcc site, as shown in FIG. 1(b); for bilayer the bottom layer remains the same while the layer above is of hcp-fcc site\cite{22,23}; in case of more than two layers of graphene, an AB stacking model is used. This configuration is shown in FIG.1(a).

Then SOC is included into calculation to determine the influence of Rashba effect on the properties of monolayer case.

\section*{\uppercase\expandafter{\romannumeral3}.\ RESULTS AND DISCUSSIONS}
The calculated band structures along $–K-\Gamma-K$ for 1-4 layers of graphene on Ni(111) are demonstrated in FIG.2. We perform projected band structure calculation, in which the radius of red/blue spots demonstrates the weight of $p_z$ orbit of top layer graphene. FIG.2(a)/2(b) display band structure near Fermi surface for majority/minority spin of monolayer case, while that of bilayer, trilayer and quadrilayer are respectively shown in FIG.2(c)/2(d), 2(e)/2(f) and 2(g)/2(h).

\begin{figure}
\includegraphics[width=7.5cm]{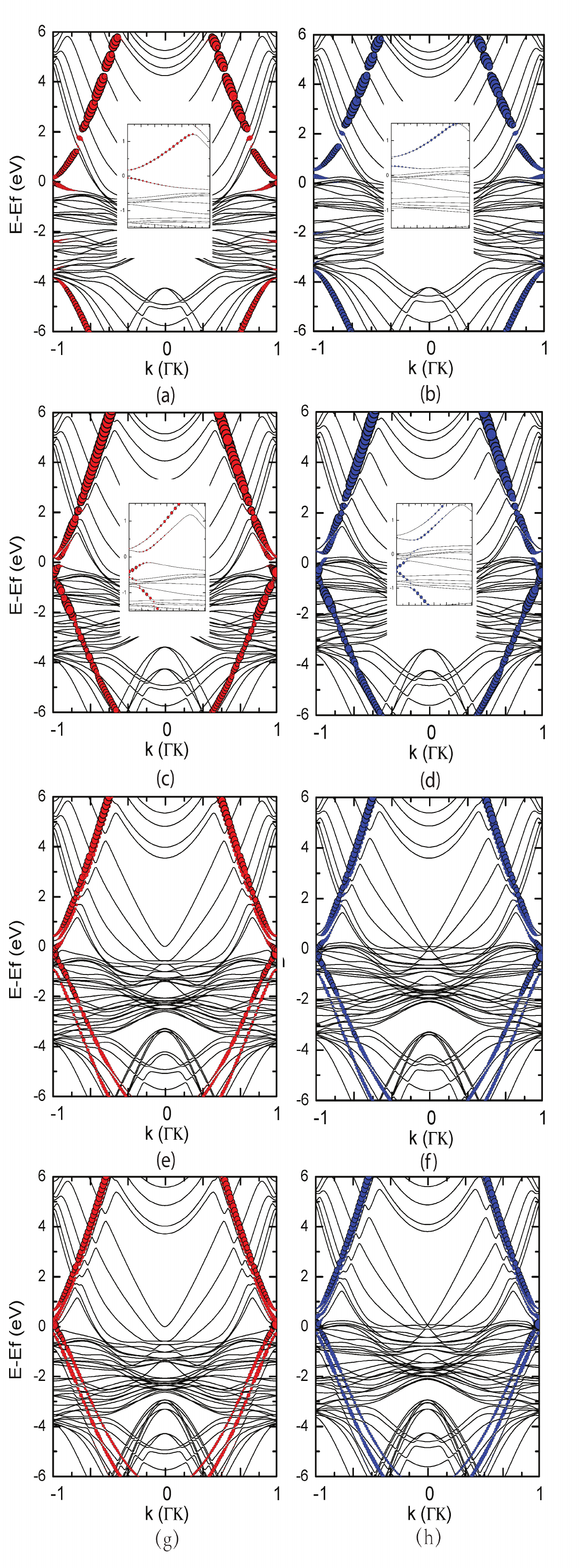}
\caption{\label{f2} (Color online) Band structure near Fermi surface for majority/minority spin of monolayer  (a) and (b), bilayer (c) and (d), trilayer (e) and (f), quadrilayer (g) and (h). The radius of red spots demonstrate the weight of $p_z$ orbit of top layer graphene. The insets show the detail near -K point and Fermi Level.}
\end{figure}

As is shown in FIG. 2(a) and 2(b), the existence of Ni substrate seriously changes the band structure of graphene compared with freestanding situation with gapless Dirac cone and linear energy dispersion. Due to the broken of the equivalence of the two sublattices of graphene caused by the substrate interaction, the Dirac Cone must be destroyed and a gap about 0.25eV opens up. The dense gray bands below the gap is those of Ni 3d orbit. The ferromagnetism of Ni leads to a 0.4eV spin gap, indicating the spin-polarized characteristics of graphene layer. The detail magnetization is displayed in FIG. 3, in which red and blue circle represents magnetism of C atoms parallel and anti-parallel to that of Ni atoms while the radius stands for the magnitude respectively. For convenience, the top layer Ni substrate is drawn as large gray circles. The graphene has an anti-ferromagnetic order, with one sublattice, namely A, under which is a first layer Ni atom, of magnetic momentum anti-parallel ($−0.01\mu B$ ) to that of Ni, and the other sublattice, namely B, of a parallel magnetism ($0.03\mu B$ ).

\begin{figure}
\includegraphics[width=7.5cm]{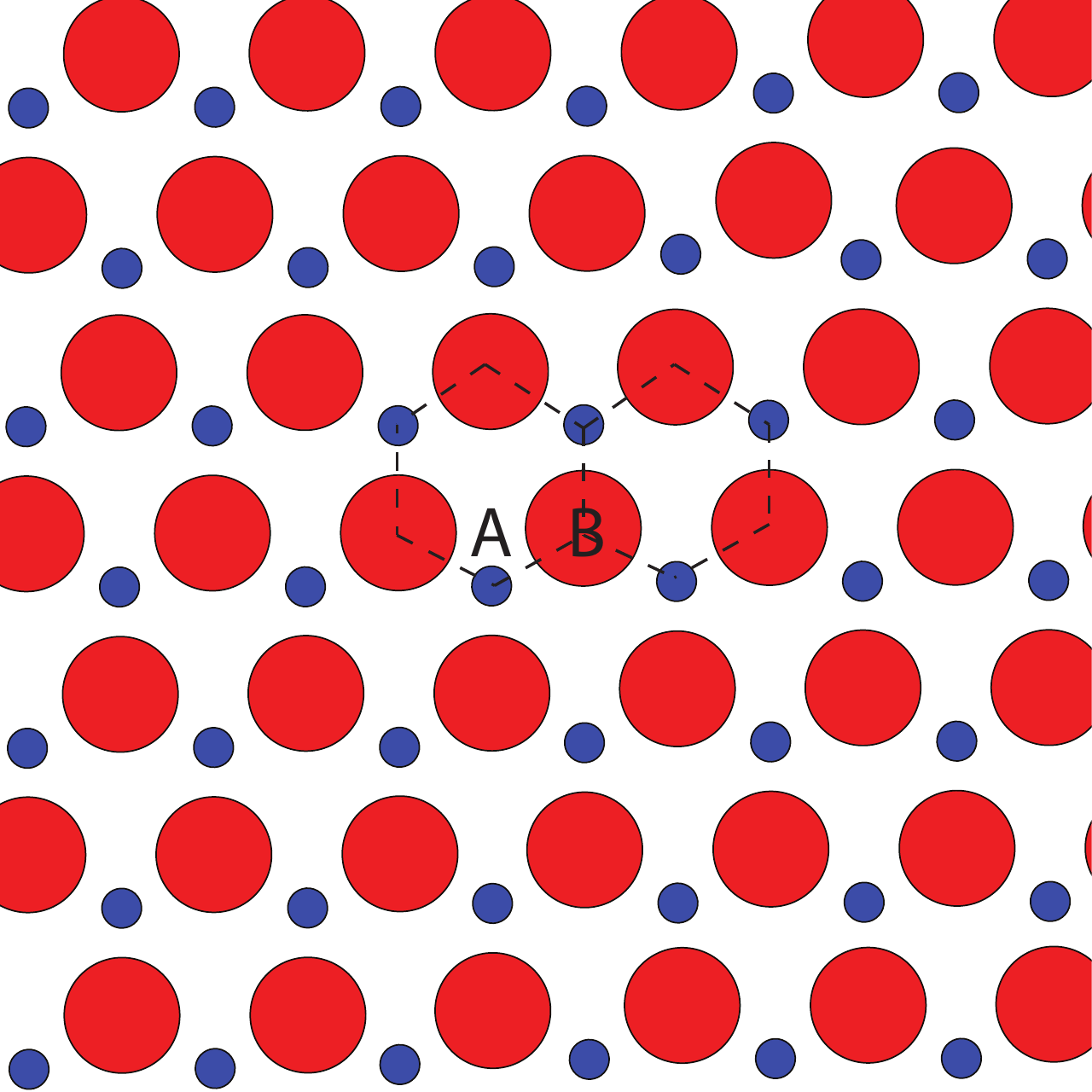}
\caption{\label{f3} (Color online) Calculated magnet moment of graphene atoms. Red/blue circle represents C atoms with magnetism parallel/anti-parallel to that of Ni atoms and the radius stands for the magnitude respectively.}
\end{figure}

For bilayer graphene, the parabolic energy dispersion is recovered since interation between graphene layers is weak. The Dirac point is lowered by about 0.4eV, suggesting strong charge transfer and vertical electric field, which can explain the small energy gap. And the spin gap is almost immeasurable, displaying the non spin-polarized character of top layer. Our calculation of the magnetization is also one magnitude lower than monolayer case. In Fig. 2(e) to 2(f), the energy of Dirac cone is at −0.2eV. In quadrilayer case, the Dirac Cone is almost at the Fermi Level, as shown in FIG. 2(g) and 2(h). This can be understood since less charge will be transferred to upper layer. And we expectedly find no difference between spins and negligible magnet moment. 

\begin{figure}[]
\includegraphics[width=8cm]{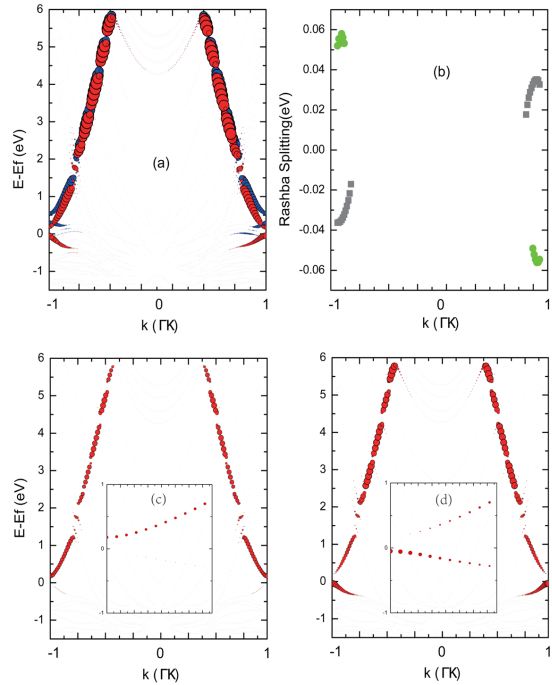}
\caption{\label{f3} (Color online) (a) Calculated band structure of monolayer graphene /Ni(111) system near Fermi level after introducing SOC. Red/blue circles stand for majority/minority spin bands, and their radius represents weight of graphene atoms' $\pi$ orbits.  (b) Rashba splitting of upper (grey square) and lower (green circle) band. (c)/(d) the contribution of $\pi$ orbit of sublattice A/B to the bands, the radius indicates the weight. The insets show the detail near -K point and Fermi Level. }
\end{figure}

Thus from the discussion above, we conclude that the Ni substrate will result in a magnetization of graphene atoms and a vertical electric field. Both effects are weaken when layer number of graphene increase.
Then Non-collinear Calculation is done for monolayer case since it has observable magnetization. We are interested in the band structure along $\Gamma-K$ near Fermi surface. The electric field E, as discussed above, is along the normal direction of Ni(111) surface, the measured wavevector $k$ is along $\Gamma-K$, to which the magnetism of the system is set perpendicular in plane. The calculated band structure of monolayer case involving SOC is shown in FIG. 4(a), in which red/blue circles stand for bands of majority/minority spin with their radius representing weight of $\pi$ orbit of graphene atoms. The $k/-k$ symmetry of bands is destroyed due to the linear dependence of Rashba splitting on wavevector. To determine the effect on spin gap of Rashba and exchange splitting individually, we do calculation of the same system both with/without SOC. With SOC, the spin gap will be $E = E_{SO} + E_{EX}$, respectively representing the contribution of SOC and exchange splitting. If without SOC, only $E_{EX}$ is included. In this way, we can show the individual Rashba splitting in FIG. 4(b). One important thing we should notice is that, for each spin in FIG. 4(a) there are two bands near the Fermi surface, one above the energy gap at K and the other below. Rashba splitting for upper/lower band is represented with grey squares/green circles in FIG. 4(b). It is critical that the Rashba splitting of upper (30meV) and lower (−60meV) bands are of different sign, which means that the energy gap of one spin at K will be enlarged while the other shrinked. This can be understood since the upper and lower bands near K and -K are mainly contributed by sublattice A and B respectively, see FIG.4(c) and 4(d). As has discussed above, the two sublattices are of opposite magnetism, which, according to Eq.(1), will result in opposite shift in spin gap. Due to the periodic boundary condition used in simulation, $E_{SO}$ decreases to zero at the boundary of BZ\cite{9}. Thus we evaluate it from near K and -K, which will be underestimated. The Rashba spin gap for upper band decrease sharply when getting away from K and -K(see the gray squares in FIG. 4(b)). This is because of the mixed contributions of both sublattices to the upper band at wavevector away from K and -K, as shown in FIG. 4(c) and 4(d). However, if we are focusing on band within 0.5eV around energy gap which mainly determains the transport property, the Rashba effect remains more than 30meV /-50meV. In cases of more than one layer of graphene, the Rashba effect on band structure of top layer also becomes inconsiderable due to the almost disappearance of magnetism.

\begin{figure}[]
\includegraphics[width=8.5cm]{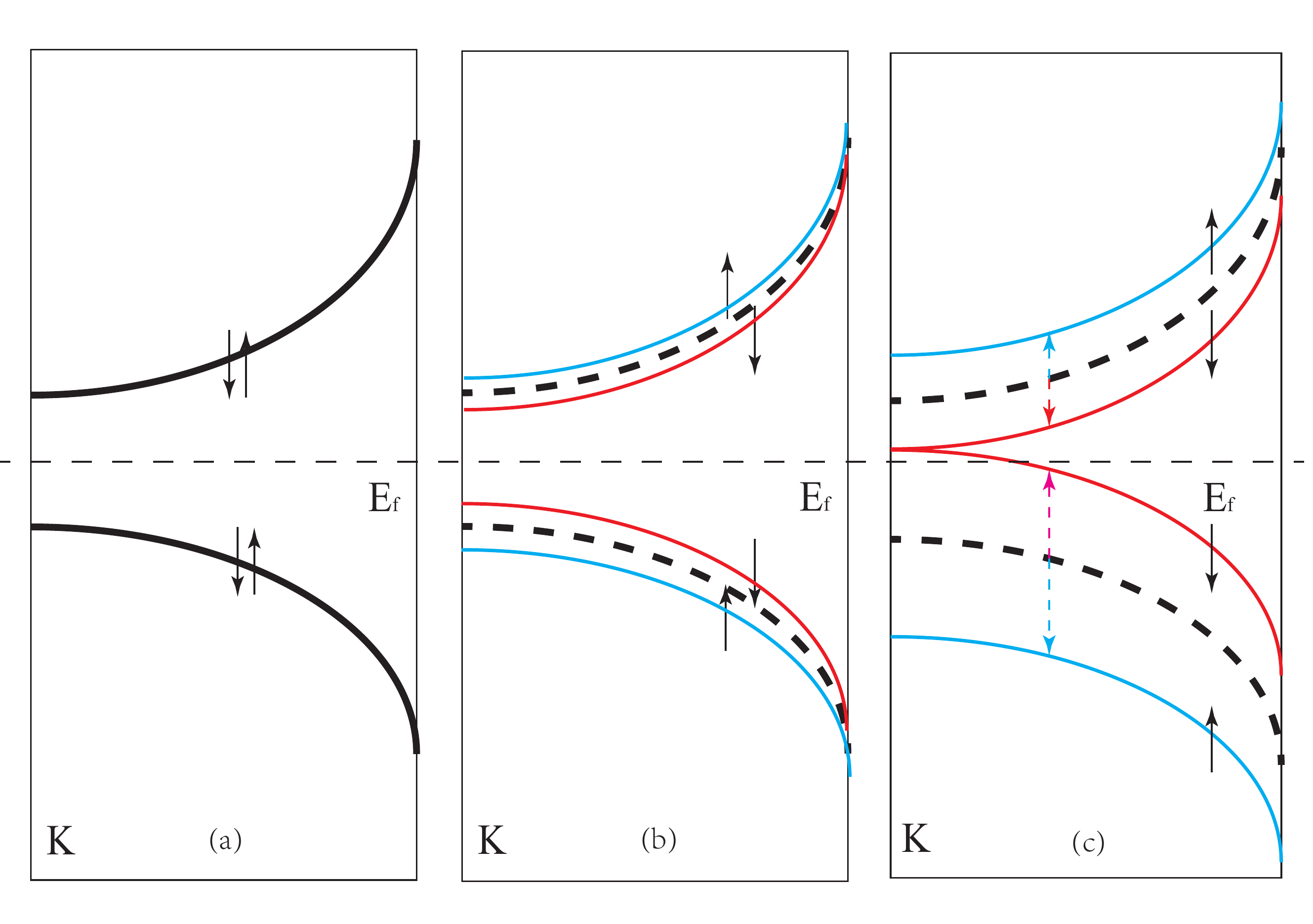}
\caption{\label{f5} (Color online) Schematic diagram to apply our method, excluding exchange splitting. (a) with zero electric field in Rashba Hamiltonian, black lines stands for bands of both spin up and spin down. (b) with the electric field in Ni(111) case, blue/red line respectively represents spin up/down. (c) with external electric field or strengthened magnetization of graphene, the Rashba effect is enlarged and the system is converted from semiconductor to half metal.}
\end{figure}

In this case, we have shown the comparatively large Rashba induced extra spin gap for bands at K near Fermi surface. In addition, the anti-ferromagnetic order and two sublattice character of graphene lead to the discrepancy of energy gap between spins. One will be enlarged for about 45meV and the other shrinked for about the same magnitude. Here the electric field in Rashba Hamiltonian is produced by the spontaneous charge transfer, which is approximately $0.4$eV/${\AA}$\cite{26}. Such effect can be tuned through extrinsic electric field perpendicular to the interface, and an electric field strong enough may completely close the gap for one spin and remarkably enlarge the gap for the other spin, which is the ideal situation in spintronics. FIG.5 is a schematic diagram in which we exclude exchange splitting and only consider Rashba effect. FIG.5(a)  is the case with zero electric field and no Rashba effect, (b) is the Ni(111) case. It may be converted from semiconductor to half metal if we futher apply external electric field or strengthen the magnetization of graphene, as shown in FIG.5(c). Another critical implication is its application in valleytronics. Due to the linear dependence of Rashba splitting on wavevector, the degeneracy of the valley degree of freedom is broken and the shrink/enlarge effect is opposite at K/-K. This indicates that under proper configuration one spin may only transport at K and the other only at -K. Thus the Rashba effect we discuss here can also be an effective valley filter.

\section*{\uppercase\expandafter{\romannumeral4}.\ CONCLUSINOS}
In summary, graphene/Ni (111) are studied considering several factors including magnetism induced by substrate and Rashba effect. At last, the Rashba effect is found to produce a 30 meV/-60meV extra spin gap on upper/lower band at K near Fermi surface, which can further adjusts the spin gap with Rashba effect. In addition, the possibility to make discrepancy between energy gaps of different spin is discovered and interpreted, which provides a completely new method to efficiently optimize graphene/Ni system for the application in spintronics. And the opposite effect at K/K' makes it also a promising valley filter. This method is tunable and can be extended to other systems with similar characters. Actually, we expect such effect in all graphene systems with different magnetization of two sublattices. Graphene on semiconductor substrate is desirable for application and SiC is a promising candidate. Another perhaps better system is carbon nanostructures without any substrate, which can display antiferromagnetic behavior due to topological frustration\cite{32}. Graphene nanoribbon is an particularly expected choice, which is under way.
\section*{ACKNOWLEDGMENT}
C.H.J. and Z.B. are supported by National Creative Fund for undergraduate. We thank Xiao Li and Junren Shi for helpful discussion.  
\bibliography{manuscript}

\begin{thebibliography}{23}
\expandafter\ifx\csname natexlab\endcsname\relax\def\natexlab#1{#1}\fi
\expandafter\ifx\csname bibnamefont\endcsname\relax
  \def\bibnamefont#1{#1}\fi
\expandafter\ifx\csname bibfnamefont\endcsname\relax
  \def\bibfnamefont#1{#1}\fi
\expandafter\ifx\csname citenamefont\endcsname\relax
  \def\citenamefont#1{#1}\fi
\expandafter\ifx\csname url\endcsname\relax
  \def\url#1{\texttt{#1}}\fi
\expandafter\ifx\csname urlprefix\endcsname\relax\def\urlprefix{URL }\fi
\providecommand{\bibinfo}[2]{#2}
\providecommand{\eprint}[2][]{\url{#2}}

\bibitem[{\citenamefont{Novoselov and et.al.}(2005)}]{28}
\bibinfo{author}{\bibfnamefont{K.}~\bibnamefont{Novoselov}} \bibnamefont{and}
  \bibinfo{author}{\bibfnamefont{A.~G.} \bibnamefont{et.al.}},
  \bibinfo{journal}{Nature} \textbf{\bibinfo{volume}{438}}
  (\bibinfo{year}{2005}).

\bibitem[{\citenamefont{Novoselov and et.al.}(2004)}]{29}
\bibinfo{author}{\bibfnamefont{K.}~\bibnamefont{Novoselov}} \bibnamefont{and}
  \bibinfo{author}{\bibfnamefont{A.~G.} \bibnamefont{et.al.}},
  \bibinfo{journal}{Science} \textbf{\bibinfo{volume}{306}},
  \bibinfo{pages}{666} (\bibinfo{year}{2004}).

\bibitem[{\citenamefont{Zhang et~al.}(2005)\citenamefont{Zhang, Tan, Stormer,
  and Kim}}]{30}
\bibinfo{author}{\bibfnamefont{Y.}~\bibnamefont{Zhang}},
  \bibinfo{author}{\bibfnamefont{Y.-W.} \bibnamefont{Tan}},
  \bibinfo{author}{\bibfnamefont{H.~L.} \bibnamefont{Stormer}},
  \bibnamefont{and} \bibinfo{author}{\bibfnamefont{P.}~\bibnamefont{Kim}},
  \bibinfo{journal}{Nature} \textbf{\bibinfo{volume}{438}}
  (\bibinfo{year}{2005}).

\bibitem[{\citenamefont{Berger and de~Heer~et.al.}(2006)}]{31}
\bibinfo{author}{\bibfnamefont{C.}~\bibnamefont{Berger}} \bibnamefont{and}
  \bibinfo{author}{\bibfnamefont{W.~A.} \bibnamefont{de~Heer~et.al.}},
  \bibinfo{journal}{Science} \textbf{\bibinfo{volume}{312}},
  \bibinfo{pages}{1191} (\bibinfo{year}{2006}).

\bibitem[{\citenamefont{Hill et~al.}(2006)\citenamefont{Hill, Geim, Novoselov,
  Schedin, and Blake}}]{10}
\bibinfo{author}{\bibfnamefont{E.~W.} \bibnamefont{Hill}},
  \bibinfo{author}{\bibfnamefont{A.~K.} \bibnamefont{Geim}},
  \bibinfo{author}{\bibfnamefont{K.}~\bibnamefont{Novoselov}},
  \bibinfo{author}{\bibfnamefont{F.}~\bibnamefont{Schedin}}, \bibnamefont{and}
  \bibinfo{author}{\bibfnamefont{P.}~\bibnamefont{Blake}},
  \bibinfo{journal}{IEEE Trans. Magn.} \textbf{\bibinfo{volume}{42}},
  \bibinfo{pages}{2694} (\bibinfo{year}{2006}).

\bibitem[{\citenamefont{Semenov et~al.}(2007)\citenamefont{Semenov, Kim, and
  Zavada}}]{20}
\bibinfo{author}{\bibfnamefont{Y.~G.} \bibnamefont{Semenov}},
  \bibinfo{author}{\bibfnamefont{K.~W.} \bibnamefont{Kim}}, \bibnamefont{and}
  \bibinfo{author}{\bibfnamefont{J.~M.} \bibnamefont{Zavada}},
  \bibinfo{journal}{Appl. Phys. Lett.} \textbf{\bibinfo{volume}{91}},
  \bibinfo{pages}{153105} (\bibinfo{year}{2007}).

\bibitem[{\citenamefont{Kane and Mele}(2005{\natexlab{a}})}]{11}
\bibinfo{author}{\bibfnamefont{C.~L.} \bibnamefont{Kane}} \bibnamefont{and}
  \bibinfo{author}{\bibfnamefont{E.~J.} \bibnamefont{Mele}},
  \bibinfo{journal}{Phys. Rev. Lett.} \textbf{\bibinfo{volume}{95}},
  \bibinfo{pages}{226801} (\bibinfo{year}{2005}{\natexlab{a}}).

\bibitem[{\citenamefont{Kane and Mele}(2005{\natexlab{b}})}]{12}
\bibinfo{author}{\bibfnamefont{C.~L.} \bibnamefont{Kane}} \bibnamefont{and}
  \bibinfo{author}{\bibfnamefont{E.~J.} \bibnamefont{Mele}},
  \bibinfo{journal}{Phys. Rev. Lett.} \textbf{\bibinfo{volume}{95}},
  \bibinfo{pages}{146802} (\bibinfo{year}{2005}{\natexlab{b}}).

\bibitem[{\citenamefont{Yao et~al.}(2007)\citenamefont{Yao, Ye, Qi, Zhang, and
  Fang}}]{13}
\bibinfo{author}{\bibfnamefont{Y.}~\bibnamefont{Yao}},
  \bibinfo{author}{\bibfnamefont{F.}~\bibnamefont{Ye}},
  \bibinfo{author}{\bibfnamefont{X.~L.} \bibnamefont{Qi}},
  \bibinfo{author}{\bibfnamefont{S.~C.} \bibnamefont{Zhang}}, \bibnamefont{and}
  \bibinfo{author}{\bibfnamefont{Z.}~\bibnamefont{Fang}},
  \bibinfo{journal}{Phys. Rev. B} \textbf{\bibinfo{volume}{75}},
  \bibinfo{pages}{041401(R)} (\bibinfo{year}{2007}).

\bibitem[{\citenamefont{van Gelderen and Smith}(2010)}]{14}
\bibinfo{author}{\bibfnamefont{R.}~\bibnamefont{van Gelderen}}
  \bibnamefont{and} \bibinfo{author}{\bibfnamefont{C.~M.} \bibnamefont{Smith}},
  \bibinfo{journal}{Phys. Rev. B} \textbf{\bibinfo{volume}{81}},
  \bibinfo{pages}{125435} (\bibinfo{year}{2010}).

\bibitem[{\citenamefont{Min et~al.}(2006)\citenamefont{Min, Hill, Sinitsyn,
  Sahu, Kleinman, and H.MacDonald}}]{15}
\bibinfo{author}{\bibfnamefont{H.}~\bibnamefont{Min}},
  \bibinfo{author}{\bibfnamefont{J.~E.} \bibnamefont{Hill}},
  \bibinfo{author}{\bibfnamefont{N.~A.} \bibnamefont{Sinitsyn}},
  \bibinfo{author}{\bibfnamefont{B.~R.} \bibnamefont{Sahu}},
  \bibinfo{author}{\bibfnamefont{L.}~\bibnamefont{Kleinman}}, \bibnamefont{and}
  \bibinfo{author}{\bibfnamefont{A.}~\bibnamefont{H.MacDonald}},
  \bibinfo{journal}{Phys. Rev. B} \textbf{\bibinfo{volume}{74}},
  \bibinfo{pages}{165310} (\bibinfo{year}{2006}).

\bibitem[{\citenamefont{Neto and Guinea}(2009)}]{16}
\bibinfo{author}{\bibfnamefont{H.~C.} \bibnamefont{Neto}} \bibnamefont{and}
  \bibinfo{author}{\bibfnamefont{F.}~\bibnamefont{Guinea}},
  \bibinfo{journal}{Phys. Rev. Lett.} \textbf{\bibinfo{volume}{103}},
  \bibinfo{pages}{026804} (\bibinfo{year}{2009}).

\bibitem[{\citenamefont{Dedkov et~al.}(2008)\citenamefont{Dedkov, Fonin,
  R{\"u}diger, and Laubschat}}]{17}
\bibinfo{author}{\bibfnamefont{Y.}~\bibnamefont{Dedkov}},
  \bibinfo{author}{\bibfnamefont{M.}~\bibnamefont{Fonin}},
  \bibinfo{author}{\bibfnamefont{U.}~\bibnamefont{R{\"u}diger}},
  \bibnamefont{and}
  \bibinfo{author}{\bibfnamefont{C.}~\bibnamefont{Laubschat}},
  \bibinfo{journal}{Phys. Rev. Lett.} \textbf{\bibinfo{volume}{100}},
  \bibinfo{pages}{107602} (\bibinfo{year}{2008}).

\bibitem[{\citenamefont{Rader et~al.}(2009)\citenamefont{Rader, Varykhalov,
  S{\'a}nchez-Barriga, Marchenko, Rybkin, and Shikin}}]{18}
\bibinfo{author}{\bibfnamefont{O.}~\bibnamefont{Rader}},
  \bibinfo{author}{\bibfnamefont{A.}~\bibnamefont{Varykhalov}},
  \bibinfo{author}{\bibfnamefont{J.}~\bibnamefont{S{\'a}nchez-Barriga}},
  \bibinfo{author}{\bibfnamefont{D.}~\bibnamefont{Marchenko}},
  \bibinfo{author}{\bibfnamefont{A.}~\bibnamefont{Rybkin}}, \bibnamefont{and}
  \bibinfo{author}{\bibfnamefont{A.~M.} \bibnamefont{Shikin}},
  \bibinfo{journal}{Phys. Rev. Lett.} \textbf{\bibinfo{volume}{102}},
  \bibinfo{pages}{057602} (\bibinfo{year}{2009}).

\bibitem[{\citenamefont{Gierz et~al.}(2010)\citenamefont{Gierz, Dil, Meier,
  Slomski, Osterwalder, Henk, Winkler, Ast, and Kern}}]{19}
\bibinfo{author}{\bibfnamefont{I.}~\bibnamefont{Gierz}},
  \bibinfo{author}{\bibfnamefont{J.~H.} \bibnamefont{Dil}},
  \bibinfo{author}{\bibfnamefont{F.}~\bibnamefont{Meier}},
  \bibinfo{author}{\bibfnamefont{B.}~\bibnamefont{Slomski}},
  \bibinfo{author}{\bibfnamefont{J.}~\bibnamefont{Osterwalder}},
  \bibinfo{author}{\bibfnamefont{J.}~\bibnamefont{Henk}},
  \bibinfo{author}{\bibfnamefont{R.}~\bibnamefont{Winkler}},
  \bibinfo{author}{\bibfnamefont{C.~R.} \bibnamefont{Ast}}, \bibnamefont{and}
  \bibinfo{author}{\bibfnamefont{K.}~\bibnamefont{Kern}},
  \bibinfo{journal}{arXiv: 1004.1573}  (\bibinfo{year}{2010}).

\bibitem[{\citenamefont{Gong et~al.}(2011)\citenamefont{Gong, Li, Yang, Gong,
  Duan, and Chu}}]{9}
\bibinfo{author}{\bibfnamefont{S.~J.} \bibnamefont{Gong}},
  \bibinfo{author}{\bibfnamefont{Z.~Y.} \bibnamefont{Li}},
  \bibinfo{author}{\bibfnamefont{Z.~Q.} \bibnamefont{Yang}},
  \bibinfo{author}{\bibfnamefont{C.}~\bibnamefont{Gong}},
  \bibinfo{author}{\bibfnamefont{C.-G.} \bibnamefont{Duan}}, \bibnamefont{and}
  \bibinfo{author}{\bibfnamefont{J.~H.} \bibnamefont{Chu}},
  \bibinfo{journal}{J. Appl. Phys.} \textbf{\bibinfo{volume}{110}},
  \bibinfo{pages}{043704} (\bibinfo{year}{2011}).

\bibitem[{\citenamefont{Kresse and Furthmuller}(1996{\natexlab{a}})}]{24}
\bibinfo{author}{\bibfnamefont{G.}~\bibnamefont{Kresse}} \bibnamefont{and}
  \bibinfo{author}{\bibfnamefont{J.}~\bibnamefont{Furthmuller}},
  \bibinfo{journal}{Comput. Mater. Sci.} \textbf{\bibinfo{volume}{6}},
  \bibinfo{pages}{15} (\bibinfo{year}{1996}{\natexlab{a}}).

\bibitem[{\citenamefont{Kresse and Furthmuller}(1996{\natexlab{b}})}]{25}
\bibinfo{author}{\bibfnamefont{G.}~\bibnamefont{Kresse}} \bibnamefont{and}
  \bibinfo{author}{\bibfnamefont{J.}~\bibnamefont{Furthmuller}},
  \bibinfo{journal}{Phys. Rev. B} \textbf{\bibinfo{volume}{54}},
  \bibinfo{pages}{11169} (\bibinfo{year}{1996}{\natexlab{b}}).

\bibitem[{\citenamefont{Hamada and Otani}(2010)}]{21}
\bibinfo{author}{\bibfnamefont{I.}~\bibnamefont{Hamada}} \bibnamefont{and}
  \bibinfo{author}{\bibfnamefont{M.}~\bibnamefont{Otani}},
  \bibinfo{journal}{PhysRevB.} \textbf{\bibinfo{volume}{82}},
  \bibinfo{pages}{153412} (\bibinfo{year}{2010}).

\bibitem[{\citenamefont{Karpan et~al.}(2007)\citenamefont{Karpan, Giovannetti,
  Khomyakov, Talanana, Starikov, Zwierzycki, van~den Brink, Brocks, and
  Kelly}}]{22}
\bibinfo{author}{\bibfnamefont{V.~M.} \bibnamefont{Karpan}},
  \bibinfo{author}{\bibfnamefont{G.}~\bibnamefont{Giovannetti}},
  \bibinfo{author}{\bibfnamefont{P.~A.} \bibnamefont{Khomyakov}},
  \bibinfo{author}{\bibfnamefont{M.}~\bibnamefont{Talanana}},
  \bibinfo{author}{\bibfnamefont{A.~A.} \bibnamefont{Starikov}},
  \bibinfo{author}{\bibfnamefont{M.}~\bibnamefont{Zwierzycki}},
  \bibinfo{author}{\bibfnamefont{J.}~\bibnamefont{van~den Brink}},
  \bibinfo{author}{\bibfnamefont{G.}~\bibnamefont{Brocks}}, \bibnamefont{and}
  \bibinfo{author}{\bibfnamefont{P.~J.} \bibnamefont{Kelly}},
  \bibinfo{journal}{Phys. Rev. Lett.} \textbf{\bibinfo{volume}{99}},
  \bibinfo{pages}{176602} (\bibinfo{year}{2007}).

\bibitem[{\citenamefont{Gamo et~al.}(1997)\citenamefont{Gamo, Nagashima,
  Wakabayashi, Terai, and Oshima}}]{23}
\bibinfo{author}{\bibfnamefont{Y.}~\bibnamefont{Gamo}},
  \bibinfo{author}{\bibfnamefont{A.}~\bibnamefont{Nagashima}},
  \bibinfo{author}{\bibfnamefont{M.}~\bibnamefont{Wakabayashi}},
  \bibinfo{author}{\bibfnamefont{M.}~\bibnamefont{Terai}}, \bibnamefont{and}
  \bibinfo{author}{\bibfnamefont{C.}~\bibnamefont{Oshima}},
  \bibinfo{journal}{Surf. Sci.} \textbf{\bibinfo{volume}{374}},
  \bibinfo{pages}{61} (\bibinfo{year}{1997}).

\bibitem[{\citenamefont{Gong et~al.}(2010)\citenamefont{Gong, Lee, Shan, M,
  Vogel, Wallace, and Cho}}]{26}
\bibinfo{author}{\bibfnamefont{C.}~\bibnamefont{Gong}},
  \bibinfo{author}{\bibfnamefont{G.}~\bibnamefont{Lee}},
  \bibinfo{author}{\bibfnamefont{B.}~\bibnamefont{Shan}},
  \bibinfo{author}{\bibfnamefont{E.}~\bibnamefont{M}},
  \bibinfo{author}{\bibnamefont{Vogel}}, \bibinfo{author}{\bibfnamefont{R.~M.}
  \bibnamefont{Wallace}}, \bibnamefont{and}
  \bibinfo{author}{\bibfnamefont{K.}~\bibnamefont{Cho}}, \bibinfo{journal}{J.
  Appl. Phys.} \textbf{\bibinfo{volume}{108}}, \bibinfo{pages}{123711}
  (\bibinfo{year}{2010}).

\bibitem[{\citenamefont{Wang et~al.}(2009)\citenamefont{Wang, Yazyev, Meng, and
  Kaxiras}}]{32}
\bibinfo{author}{\bibfnamefont{W.~L.} \bibnamefont{Wang}},
  \bibinfo{author}{\bibfnamefont{O.~V.} \bibnamefont{Yazyev}},
  \bibinfo{author}{\bibfnamefont{S.}~\bibnamefont{Meng}}, \bibnamefont{and}
  \bibinfo{author}{\bibfnamefont{E.}~\bibnamefont{Kaxiras}},
  \bibinfo{journal}{Phys. Rev. Lett.} \textbf{\bibinfo{volume}{102}}
  (\bibinfo{year}{2009}).

\end{thebibliography}
\end{document}